\documentclass[%
 reprint,
 amsmath,amssymb,
 aps,
showkeys,
]{revtex4-2}
\usepackage{graphicx}% Include figure files
\usepackage{dcolumn}% Align table columns on decimal point
\usepackage{xcolor}
\usepackage{bm}% bold math
\usepackage{lipsum}
\usepackage{physics}

\makeatletter
\renewcommand\@seccntformat[1]{}
\renewcommand\subsection{\@startsection{subsection}{2}{0pt}%
  {-3.25ex \@plus -1ex \@minus -.2ex}%
  {1.5ex \@plus .2ex}%
  {\centering\normalfont\normalsize\bfseries}}
\makeatother

\begin{document}

%\preprint{APS/123-QED}

\title{Multimodal oscillator networks learn to solve a classification problem}% Force line breaks with \\

\author{Daan de Bos}
\author{Marc Serra-Garcia}%
\affiliation{%
AMOLF\\
 Science Park 104, 1098XG Amsterdam, The Netherlands
}%

\date{\today}% It is always \today, today,
%  Physical information processors, such as photonic Ising samplers, memristor crossbar arrays and Hopfield networks, are already capable of solving optimization problems, performing inference and recalling associative memories through their intrinsic physical evolution. However, their internal parameters are typically fixed or adjusted through external control circuits.
\begin{abstract}
We numerically demonstrate a network of coupled oscillators that can learn to solve a classification task from a set of examples---performing  both training and inference through the nonlinear evolution of the system. We accomplish this by combining three key elements to achieve learning: A long-term memory that stores learned responses, analogous to the synapses in biological brains; a short-term memory that stores the neural activations, similar to the firing patterns of neurons; and an evolution law that updates the synapses in response to novel examples, inspired by synaptic plasticity. Achieving all three elements in wave-based information processors such as metamaterials is a significant challenge. Here, we solve it by leveraging the material multistability to implement long-term memory, and harnessing symmetries and thermal noise to realize the learning rule. Our analysis reveals that the learning mechanism, although inspired by synaptic plasticity, also shares parallelisms with bacterial evolution strategies, where mutation rates increase in the presence of noxious stimuli.
\end{abstract}

\maketitle

%\tableofcontents

\section*{Introduction}\label{Intro}
% This should be a bit more explained while staying in 1-2 sentences , e.g. While physical computing has achieved remarkable successes in diverse fields from image and sound recognition, embodied intelligence in soft robots, differential equation solving and linear algebra. 
% Perhaps a description of what physical computing is would also help: Physical computing aims to leverage phenomena such as noise, symmetry, interference, parametric phenomena, topology, etc to direclty evaluate functions and implement algorithms.  It has achieved remarkable successes in A, B and C... [note that it is still only 2 sentences! ]
%The description of learning should also be clarified, like the realization of 'physical learning systems, physical computers that can autonomously adapt in response to feedback and examples, is still in its infancy.

% Alternative:
% Metamaterials are an excellent platform to compute. Metamaterial systems have been shown to be capable of computing arithmetics, derivatives and integrals, equation solving, speech and image recogntion among others. Yet materials---even tunable ones---are typically designed though a top-down, centralized process. In contrast. This is challenging because the material needs to perform two tasks at once, in one hand, inference, in the other hand, parameter updates. Such systems have only worked by 

Wave-based computing offers the tantalizing prospect of massive parallelism~\cite{rajabalipanah2022parallel}, speed-of-light operation~\cite{hu2024diffractive} and low energy consumption~\cite{wang2022optical}. Applications span a broad range of problems, including evaluating arithmetic operations~\cite{silva2014performing}, computing derivatives~\cite{metasurfaceComputingPolman}, solving integral equations~\cite{cordaro2023solving}, performing voice recognition~\cite{hughes2019wave, dubvcek2024sensor}, image classification~\cite{ashtiani2022chip, lin2018all}, and generating pictures through reverse diffusion processes~\cite{oguz2024optical, bosch2025local}. Traditionally, metamaterials that compute with waves are designed for fixed tasks via top-down algorithms such as adjoint optimization~\cite{luce2024merging, bordiga2024automated} or combinatorial methods~\cite{cui2014coding}. This top-down approach contrasts with the continuous learning process that is characteristic of living brains, in which connections between decentralized neurons are autonomously formed using local information~\cite{hebb1949organization, Hebbian, wright2025distinct}.  This decentralized learning process confers a resilience and adaptability that artificial systems lack, and reduces training costs.

Physical learning is an emerging field that aims at bringing the adaptability of biological organisms to artificial materials, by creating synthetic systems that learn from examples. A key strategy in physical learning consists in updating the model parameters according to a learning rule. These local learning rules are inspired by the synaptic plasticity mechanisms in neurons, with Hebbian learning~\cite{hebb1949organization, Hebbian}---the notion that neurons form connections based on correlations between firing patterns---being one of the foremost examples. Yet learning strategies also include many recent additions, such as equilibrium propagation~\cite{scellier2017equilibrium}, the forward-forward algorithm~\cite{momeni2023backpropagation} and Hamiltonian echo backpropagation~\cite{HEB}. While much of the work has focused on developing and analyzing learning rules in abstract or algorithmic terms, an equally important frontier lies in the physical realization of systems that embody these rules. In this context, self-learning systems have been realized by externally updating the system parameters according to a learning rule, either manually~\cite{altman2024experimental} or using electronic circuits~\cite{labDem}. 

%to the plasticity rules in neurons, including the contrastive---where two copies of the system are compared, and the Hebbian---where connections are strengthened between neurons that spike simultaneously.

\begin{figure*}[!]
\includegraphics[width = \textwidth]{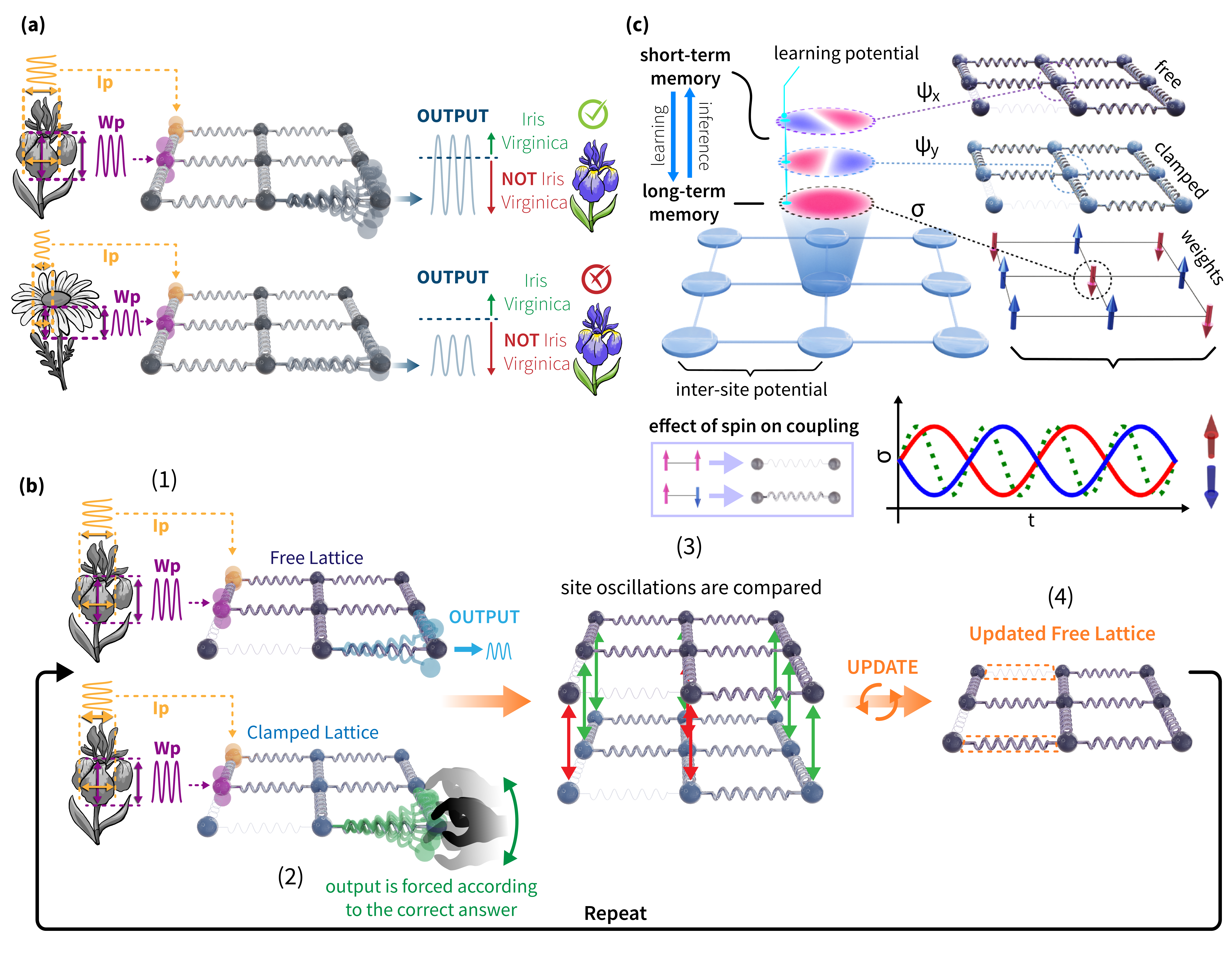}
\caption{\label{fig:epsart} Self-learning metamaterial \textbf{(a)} During inference, the metamaterial acts as a mass-spring network. Features are encoded in the excitation amplitude at designated input sites (yellow and purple spheres). If the output amplitude exceeds a threshold, an object is recognized. \textbf{(b)} For the material to learn according to a contrastive learning rule, an identical copy to the lattice in (a) is constructed. The original lattice is operated normally, letting the output site vibrate freely, and is referred to as the free lattice. In the copy, the output is clamped to vibrate at the correct amplitude. At every learning iteration, the coupling springs are updated depending on the difference between the vibrating amplitude of the clamped and free lattices. \textbf{(c)} In the proposed metamaterial, the masses of the free and clamped lattices are mapped to normal modes of each metamaterial site. We refer to these as $\psi_x$ and $\psi_y$ respectively (collectively referred as computational degrees of freedom). The coupling springs are encoded in an additional mode, that is represented by $\sigma$. This mode is parametrically driven at twice its natural frequency (dotted green line) and consequently has two stable phases of oscillation (represented as spin states with blue and red arrows). As a consequence of a nonlinear interaction potential, the effective springs connecting the computational modes depend on the weight state---taking a  weak value when neighboring strings are aligned, and a strong value when neighboring weights are oscillating in opposite phases. The learning potential (blue line connecting the modes) causes the weight state to change when clamped and free copies disagree, implementing a contrastive learning rule.}
\vspace{-10pt}
\end{figure*}

%A multifield coherent Ising machine is a network of multi-modal oscillators. It simultaneously solves two problems, training and inference. represented by a scalar field $\sigma$ interacting with a spinor field $\vec{\psi}$. At every site $i$, the model has three degrees of freedom $\psi_{x,i}$, $\psi_{y,i}$ and $\sigma_{i}$, coupled via nonlinear terms $H_L$ and $H_I$. The field $\vec{\psi}$ encodes the computational degrees of freedom (neural activations). $\vec{\psi}$ is composed of two identical sub-systems $\psi_x$ and $\psi_y$ (mass-spring lattices). During training, the output degree of freedom of the sub-system $\psi_y$ is clamped to a target value ($\psi_x$ is left free). The modes $\sigma_i$ are parametrically pumped (dotted green line) into a bistable self-oscillation regime, modeled by a spin (up or down depending on the phase of oscillation). This results in an Ising-like system that provides a long-term memory and stores the learned weights. The nonlinear interaction term $H_I$ modulates the dynamics of the computational field $\vec{\psi}$ according to the weights $\sigma$ (resulting in strong springs between antiferromagnetically-aligned sites). The term $H_L$ updates the weights $\sigma$ in response to a difference between the free ($\psi_x$) and clamped ($\psi_y$) computational sub-systems. In a material, the field $\vec{\psi}$ can be realized by mapping the two spin components to a pair of doubly-degenerate localized orbitals. }

An open problem in physical learning is figuring out how to build physical systems where the learning occurs through the natural evolution of the material, instead of relying on manual updates or external electronic circuitry. Such material could potentially extend the parallelism and speed of wave computing to the learning process, enabling the use of physical phenomena to train machine learning models. However, realizing self-learning in metamaterials is a significant challenge. This is because a self-learning material is tasked with solving two simultaneous problems~\cite{stern2025physical}: On one hand, the material performs a direct problem on the input signals. This direct problem can be, for example, solving a classification task. On the other hand, the material is also performing an inverse problem, autonomously adapting its parameters to improve its fitness in response to new examples.

 \section*{Results}
 
% \subsection{Self-learning metamaterial model}
% metamaterial model that self-learns. 
 %The metamaterial consists of a network of multimodal resonators, and learns to solve a classification task from examples. 
 
 In this work, we introduce a metamaterial model that learns from examples. The metamaterial, consisting of a  network of coupled multimodal resonators, simultaneously solves both learning and inference problems using wave physics.  When solving a classification task, the proposed metamaterial can be seen as a square lattice of masses and springs. The input is applied as a harmonic force in one or several sites. In a flower classification problem, this input may consist of the geometric features of the flower, such as petal width and petal length---with higher amplitudes corresponding to larger feature values. The output of the computation is the vibration amplitude of a designated output site (Fig.~\ref{fig:epsart}a). Although spring networks (and mathematically analogous resistor networks) are relatively simple physical systems, they are able to solve classification~\cite{li2024training} and allostery~\cite{rocks2017designing} tasks; when they are augmented with nonlinearity, they are in fact computationally universal~\cite{scellier2025universal}. The learning process is based on the contrastive learning rule~\cite{CLflow}. In contrastive learning, two copies of the system are constructed. Both copies are excited with the same input. During training, the output of one of the copies is clamped to the target value, while the other copy is left free. Then, the parameters (springs) are updated proportionally to the difference between vibration amplitudes in the clamped and free copies (Fig.~\ref{fig:epsart}b). This learning process can be intuitively understood as changing the system parameters until the free copy---which has not been exposed to the correct answer---learns to mimic the copy that is clamped to the correct result.

The requirement for two copies of the system has been an obstacle to the realization of contrastive learning schemes~\cite{anisetti2024frequency}. Here, we overcome this challenge by relying on the crystalline symmetries of the metamaterial (Fig.~\ref{fig:epsart}c). When a lattice exhibits a symmetry, such as the four-fold rotational symmetry $C_4$ present in our case, eigenmodes corresponding to two-dimensional irreducible representations of the symmetry automatically appear in pairs~\cite{dresselhaus2007group}. Around the frequency of these eigenmodes, the system effectively behaves as two separate, identical oscillator networks. These two networks can be addressed independently by controlling the location at which excitation and clamping are applied. We will use these two symmetry-protected, identical copies to construct the free and clamped copies required by the contrastive learning rule. We will denote the modal displacements at site $i$ by ${\psi_i}_x$ and ${\psi_i}_y$ respectively, and refer to both of them together as the computational field $\vec{\psi}$.

In self-learning mass-spring networks, the values of the springs connecting the computational degrees of freedom are also dynamical variables. To allow for the spring constants to vary, we encode them in an additional site eigenmode, that we refer to as the weight field $\sigma_i$. While the computational field $\vec{\psi}$ is different for every input sample---for example, for every flower being classified---the weight field $\sigma_i$ accumulates all prior learned experience of the metamaterial. Thus, it requires a much longer memory time scale. We realize this long-term memory by parametrically driving the mode, modulating its local stiffness at two times the modal resonance frequency. When a system is parametrically driven strongly enough, the system undergoes a bifurcation from which two stable, self-oscillating solutions emerge. These two stable solutions---characterized by opposite phases of oscillation---can be extremely long-lived, with transitions driven by thermal noise. Since the two stable states can be modeled by a spin-like binary variable, the weight field $\sigma$ can be thought of as emulating an Ising model. Such emulators are commonly referred to as coherent Ising machines.

The self-learning metamaterial can be seen as a coherent Ising machine $\sigma_i$ interacting with the computational field $\vec{\psi}_i$. Thus, we refer to it as a multifield coherent Ising machine. Throughout the rest of this section we will discuss how the learning rule emerges from nonlinear interactions between computational and weight fields. The nonlinearity plays a double role. First, it makes the springs that connect computational degrees of freedom dependent on the weight field $\sigma$. Second, it modulates the long-term memory of the weight field $\sigma$ depending on the difference between clamped and free computational fields---causing the system to \lq forget' the weights on sites where clamped and free copies disagree. We will finally show how these two interactions together allow the system to learn a classification problem.

The equations of motion of the material, describing the evolution of the fields $\vec{\psi}$ and $\sigma$ at each site $i$ are:

\begin{align}
&\ddot{\vec{\psi}}_{i} + \frac{\omega_c}{Q_c}\dot{\vec{\psi}}_{i} + w_c^2\vec{\psi}_{i} + \nabla_{\vec{\psi}_i} H_l = \vec{\xi}_{\psi, i};\\
&\ddot{\sigma}_i + \frac{\omega_l}{Q_l}\dot{\sigma}_i + \omega_l^2(1 + \alpha\sin(2\omega_l t))\sigma_i + \epsilon\sigma_i^3 +\nabla_{\sigma_i} H_l \ 
= \xi_{\sigma, i}. \label{eqn:Ising}
\end{align}

We set the frequency of the computational and learning degrees of freedom to $\omega_c^2=0.5$ and $\omega_l^2=1$ respectively (the subindex is used to distinguish between computational and learning modes). The corresponding quality factors are set to $Q_c=4000$ and $Q_l=40$, while the parametric driving strength---how much the resonator frequency is externally varied over time---is set to $\alpha_l=0.1$. The nonlinear Duffing coefficient plays a critical role in preventing the amplitude of the mode $\sigma_i$ to grow without bounds, and is set to $\epsilon=0.005$. The terms $\xi$ represent the thermal noise exerted by the environment on the system. They take the usual form of a set of independent, Gaussian distributed random forces with autocorrelation $\langle\xi_{l/c}(t_0)\xi_{l/c}(t_1)\rangle = \sqrt{2k_BTb_{c/l}}\delta(t_1-t_0)$. Interactions between degrees of freedom are captured by the learning potential $H_l=H_I+H_L$, where the local part $H_L$ couples the fields locally at every site of the lattice---updating the long-term memory according to the learning rule---and the interaction part $H_I$ couples each site with its nearest neighbors---ensuring that the field $\vec{\psi}$ processes information according to the weights stored in the $\sigma$ field.

\begin{figure}[t!]
\includegraphics[width = \columnwidth]{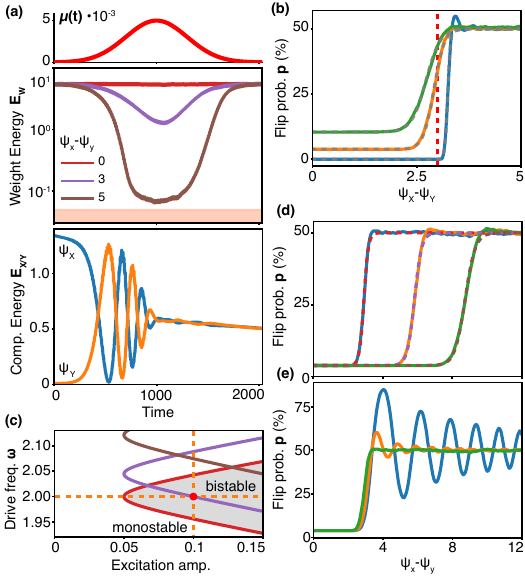}
\caption{\label{fig:flip} Single-site learning dynamics. \textbf{(a)} Evolution of the energies in the weight ($\sigma$, middle panel) and computational ($\psi_x$, $\psi_y$, bottom panel) resonators, as a learning pulse is applied (top panel). \textbf{(b)} Probability of flip during a learning protocol as a function of $\psi_X-\psi_Y$ for different temperatures, corresponding to $k_BT$ values of $0.001$ (blue), $0.05$ (orange) and $0.1$ (green). \textbf{(c)} Parametric self-oscillation region (shaded grey area) of the weight degree of freedom $\sigma$; in the shaded region, the weight is bistable and thus has long term memory.  The excitation conditions are shown as dashed orange lines and red dot. When $\psi_X-\psi_Y$ is not zero, applying the learning potential shifts the self-oscillation region. The value of $\psi_x-\psi_y$ where $\sigma$ goes out of parametric resonance is shown as a red dashed line in panel b. \textbf{(d)} Probability of spin flip as a function of $\psi_x-\psi_y$, for learning potentials $\mu_m=\mu_{m,0}/n^2$ with $n=1$ (blue), $n=2$ (orange) and $n=3$ (green), illustrating how the step location can be shifted by setting the potential. \textbf{(e)} Probability of spin flip as a function of the protocol durations $\Delta_t=50$ (blue), $\Delta_t=100$ (orange), and $\Delta_t=300$ (green), showing the emergence of probability oscillations at short learning pulses. The dashed curves in panels \textbf{(c)} and \textbf{(d)} show the fit with Eq.~\ref{eqn:switchprobability}. Throughout the figure, the parameters are $2\omega_l=2$, $\omega_l^2\alpha=0.1$, $\mu_{m,0}=0.005$, and the step duration is  $\Delta_T=300$ unless stated otherwise. } 
\vspace{-15pt}
\end{figure}

\subsection*{Single-site learning dynamics}\label{SysDesc}
In the proposed metamaterial, the learning rule arises due to local interactions at each site $i$. These interactions update the weight field ($\sigma$) when the free ($\psi_x$) and clamped ($\psi_y$) sub-systems disagree. In traditional contrastive learning, the weights are continuous variables, adjusted in proportion to the difference between clamped and free configurations $\psi_x-\psi_y$. However, in our model, every site of the weight field has only two stable phases of oscillation and, therefore, cannot be varied continuously. We adapt to this difference by implementing a probabilistic update process, where the probability that a weight $\sigma_i$ flips increases with increasing $|\psi_x-\psi_y|$. To realize this probabilistic bit flipping, we rely on the fact that the memory of a parametric oscillator---the average time it takes for the state to flip randomly due to thermal noise---is highly dependent on the detuning~\cite{dykman1998fluctuational}---the difference between its parametric excitation frequency and the optimal parametric resonance condition. We can thus realize probabilistic contrastive learning by applying a potential that shifts the natural frequency of $\sigma_i$ as a function of $\psi_y-\psi_x$, causing the system to be detuned. 

Such selective frequency shift can be accomplished with a  cross-Kerr interaction. Cross-Kerr interactions shift the frequency of one mode as a function of the energy in another mode, and take the form
%\textcolor{red}{Here we explain that the local term is responsible for the learning rule. We explain that the learning rule is continuous but the system deterministic, and that we solve this by going probabilistic}\textcolor{red}{We explain how we achieve this by controlling the memory of the resonator.}
%\textcolor{red}{We also explain that the interaction could be realized experimentally}

%Our learning rule boils down to a conditional spin forgetting mechanism: the phase of a spin $\sigma_i$ is randomized with a probability conditional on the difference between the clamped and free configurations at the site, thereby inducing spin flips in sites where free and clamped are misaligned. This mechanism arises from the following nonlinear, local interaction term acting on each site $i$,
\begin{equation}
H_{L,i}=\mu(t)\psi_{x-y,i}^2\sigma_i^2=\mu(t)\left( \psi_{x,i}-\psi_{y,i} \right)^2\sigma^2_i .
\label{eqn:localterm}
\end{equation}
Here, to turn a cross-Kerr interaction into a difference-sensitive detuning, we used the rotational symmetry of the site. Since the sites are symmetric, any combination between the eigenmodes $\psi_x$ and $\psi_y$ is also an eigenmode. Thus, we can look at the rotated set of modes $(\psi_{x+y}, \psi_{x-y})=(\psi_x+\psi_y, \psi_{x}-\psi_y)$. In this new basis, the difference between modes $\psi_{x}-\psi_y$ can be seen as its own mode $\psi_{x-y}$. In an experimental setting, such a rotated cross-Kerr interaction could be realized by introducing a non-linear spring in a region of the resonator where the two modes $\psi_x$ and $\psi_y$ have opposite signs.

The potential in Eq.~\ref{eqn:localterm} breaks the $C_4$ rotational symmetry of the resonator; when this term is active, the system can no longer be seen as two separate copies. We overcome this by setting the strength $\mu(t)$ to zero during computation, preserving the two-copy picture. To initiate a weight update, we briefly turn on the cross-Kerr term, by increasing $\mu(t)$ according to a Gaussian pulse with peak value $\mu_m$ and a width $\Delta_t$,
\begin{equation}
\mu(t)=\mu_m \exp{\left(-\frac{(t-t_0)^2}{2\Delta_t^2}\right)}.
\label{eqn:mut}
\end{equation}
Although here we chose to apply the learning potential $\mu(t)$ according to a Gaussian pulse, its specific shape is not critical. In the Appendix, we show that following a square pulse results in similar dynamics. Because $\mu(t)$ is independent of the state of the system, it is the same for every resonator and does not require a controller; its role is that of a system clock, indicating when learning shall occur.

Figure~\ref{fig:flip}a shows the evolution of the energies in the modes $\sigma$, $\psi_x$ and $\psi_y$ during a weight update. The resonator is initialized with specific values of the fields $\vec{\psi}$ and $\sigma$. During operation, the initial condition for $\vec{\psi}$ is set by the interaction with the neighbors, as will be discussed later in the paper. Our simulations reveal that, when the learning potential is applied, the oscillation of the weight degree of freedom ($\sigma)$ decreases with increasing difference between initial conditions of $\psi_x$, $\psi_y$. This is a consequence of the increased detuning of $\sigma$, which drives the mode away from resonance. The lower amplitude of the detuned mode makes thermally induced flips more likely, thus reducing the memory of the system. When the learning potential is active, we also observe a periodic exchange of energy between free and clamped modes. This is an expected consequence of $H_L$ breaking the fourfold rotational symmetry---introducing a difference between the $x+y$ and $x-y$ directions and thus distorting the picture where the clamped and free copies are independent.

The difference between clamped and free oscillations affects the probability of flipping a weight in a step-like fashion (Fig.~\ref{fig:flip}b).  This step-like response arises because $\sigma_i$ only self-oscillates in a specific region of parametric excitation frequencies and amplitudes (Fig.~\ref{fig:flip}c). When the difference between clamped and free oscillators is large enough, the nonlinear interaction (Eq.~\ref{eqn:localterm}) shifts this region outside of the excitation conditions. As a consequence, the system ceases to self-oscillate and forgets its state. At low temperature, this phenomenon is abrupt---as long as the excitation remains within the self-oscillation region, the system remembers its state. In contrast, when the temperature is finite, random thermal flips start to occur at finite amplitudes, smoothing out the response (Fig.~\ref{fig:flip}b). The difference $\psi_x-\psi_y$ at which the forgetting transition occurs is a tunable parameter. We can control it by adjusting the strength of the interaction $\mu_m$ (Fig.~\ref{fig:flip}d). This state-dependent switching relies on the ratio between the quality factors of the computational and learning degrees of freedom. Because outside parametric resonance the computational degree of freedom has a much longer memory lifetime than the weight degree of freedom, the difference between clamped and free copies remains throughout the update protocol. In a practical scenario, this difference could be implemented by controlling the coupling between each mode and the environment, or via external cooling signals to speed up the learning dynamics.

\begin{figure}[t!]
\includegraphics[width = \columnwidth]{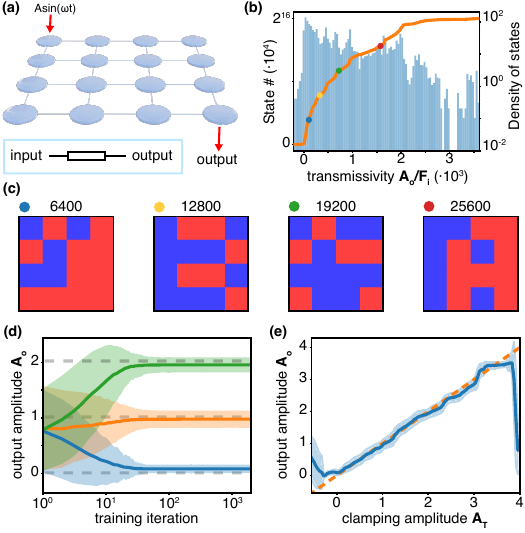}
\caption{Site-site interactions in a lattice \textbf{(a)} Example metamaterial, consisting of a 4x4 lattice. A harmonic force with frequency $\omega_c$ and amplitude $F_i$ is applied to both clamped and free degrees of freedom at the input site, situated at the top-left corner. The output amplitude $A_o$ is measured at the bottom-right corner.  \textbf{(b)}  Cumulative density of states (orange) and density of states (blue) as a function of the transmissivity, computed via the effective mass-spring model. The dots correspond to simulations of the transmissivity  using the full nonlinear system (Eq.~\ref{eqn:Ising}) with an excitation force of $10^{-4}$.  \textbf{(c)} Weight configurations corresponding to the dots in (b) \textbf{(d)} Evolution of the transmissivity as a function of training iteration for target values $A_T$ of $0$, $1$ and $2$ (dashed line). \textbf{(e)} Output amplitude $A_o$ as a  function of the clamping amplitude, after 200 learning iterations, starting from a random configuration. The shaded area represents one standard deviation. The dashed orange line corresponds to an ideal learning response. We observe that, for transmissivity values for which a weight configuration exists, the transmitted amplitude approximately converges to the clamping amplitude. Panels (d) and (e) have been computed by averaging 2000 training runs, with the shaded area representing the standard deviation. }
\label{fig:lattice} 
\vspace{-15  pt}
\end{figure}

We empirically observe that the probability of transitions  $p$ closely follows (Figs.~\ref{fig:flip}b,d) a hyperbolic tangent relation, 
\begin{equation}
p=p_0+\frac{1-2p_0}{4}\left[1+\text{tanh}\left(\beta(|\psi_{x-y}|-\psi_T)\right)\right],
\label{eqn:switchprobability}
\end{equation}
in which $\psi_T$ is the threshold for bit flipping, $\beta$ determines the slope of the flipping transition, and $p_0$ is the probability of spontaneous thermal flipping in absence of contrast between clamped and free configurations. This is unsurprising, because our weight field behaves as an effective spin system, and hyperbolic tangents commonly arise in spin-flip probabilities due to detailed balance considerations, as exemplified paradigmatically by Glauber dynamics~\cite{glauber1963time}. Later, Eq.~\ref{eqn:switchprobability} will allow us to predict the probability of bit flips in large lattices without having to simulate expensive stochastic differential equations for each site. Remarkably, the sigmoidal description breaks down for shorter pulse widths $\Delta_t$. In these cases, the oscillation of $\sigma$ does not have time to decay below the thermal noise, thus forgetting its state. The remaining amplitude is phase-shifted by the learning potential, giving rise to oscillations of the bit-flip probability as a function of $\psi_x-\psi_y$ (Fig.~\ref{fig:flip}e) when the accumulated shift is a multiple of $\pi$. In the rest of this work, we will operate in the sigmoidal regime---corresponding to long learning pulses---to achieve a monotonic relation between weight flip probability and clamped-free difference.

\subsection*{Site-site interactions}\label{Master}

%In the previous section, we have shown how the metamaterial updates its weight field when the clamped and free copies of the system disagree, following a probabilistic version of the contrastive learning rule. However, until now, we have treated the field has a simple additional degree of freedom, that has no effect on the computations that the metamaterial performs. For the material to perform computations according to its learned capabilities. Our metamaterial achieves this though a nonlinear  site-site interaction of the form

%Equally important is that the weight field conditions how computations are performed. Our metamaterial accomplishes this through a nonlinear interaction between sites. The interaction has the form:

%Where X and Y... Such nonlinear interaction can be realized, for example, by a set of cubic spring (See Appendix). To prevent the from altering the spin flip dynamics, we make the force time-dependent. However, in contrast with the previous section, the site-site interaction is on during inference, and we only set it to zero briefly at every learning iteration to induce a weight update according to the free-site dynamics.

In machine learning models, the role of the weights is to determine what computation is carried out, by setting the strength of the coupling between neurons. Consequently, in a  self-learning physical system, storing weight and computational degrees of freedom separately is not sufficient; we must also ensure that the weights influence the computation that is being performed. In our metamaterial, this influence is established through a nonlinear interaction $H_I$ acting between each pair of degrees of nearest-neighbor sites $i$, $j$. It is governed by an energy of the form

\begin{equation}
H_I=-\lambda(t)\left[c_0 + (\sigma_i-\sigma_j)^2 \right]\vec{\psi}_i^T \vec{\psi}_j,
\label{eqn:interactionpotential}
\end{equation}
This term generates the effective coupling springs that connect neighboring computational degrees of freedom ($\psi_{x,i}\leftrightarrow\psi_{x,j}$ and $\psi_{y,i}\leftrightarrow\psi_{y,j}$). Depending on the phase of the weight degrees of freedom, these coupling springs can take two values (Fig.~\ref{fig:epsart})---determined by $\lambda(t)=\lambda_m$ and $c_0$; if the weight field has the same phase of oscillation in both sites, the spring constant will be $k_{low}=\lambda_mc_0$, while if weights oscillate at opposing phases, the spring constant will be $k_{high}=\lambda_m(c_0+4<\sigma^2>)$. This effective spring model is valid as long as the computational amplitude remains small and the potential in Eq.~\ref{eqn:interactionpotential} does not alter the weight amplitude. Similarly to $\mu(t)$,  $\lambda(t)$ is a state-independent signal that is identical for all resonators, acting as a system clock. 

Although each individual weight $\sigma_i$ is binary, in metamaterials composed of multiple sites (Fig.~\ref{fig:lattice}a) the transmitted energy between input and output can take many different values, depending on the global weight configuration. This can be seen in the density of states, which quantifies the number of weight configurations that result in a lattice transmissivity around a specific value (Fig.~\ref{fig:lattice}b,c). This distribution depends on lattice size and on the $k_{high}/k_{low}$ ratio, which is chosen (approximately, by trial and error) to have the transmissivities spread out as uniformly as possible.  Because the lattice can be trained to approximate continuous transmissivities, as long as they fall within the range of expressible values, we treat this element as a single programmable weight---which we refer to as a learnistor. Networks of trainable weights, combined with fixed nonlinear activation functions, are computationally universal~\cite{cybenko1989approximation}; thus, we hypothesize that such trainable elements can form building blocks for more advanced learning architectures.

To perform inference with the metamaterial, we encode the input in the amplitude of a harmonic force, applied to both computational degrees of freedom, $\psi_{x, input}$ and $\psi_{y, input}$, at the input site. We let the lattice reach a steady-state, and read the result in the amplitude of the output computational degree of freedom, $\psi_{x, output}$. During this process, we keep $\lambda(t)$ at its inference value $\lambda_m$ and $\mu(t)$ at zero. To perform a training iteration, in addition to the input forces, we also clamp the output site $\psi_{y, output}$ to the desired value. Then, once the lattice reaches a steady-state, we perform a weight update by increasing $\mu(t)$ according to the protocol discussed in the previous section. During the weight update, we momentarily set $\lambda(t)=0$, turning the lattice into a set of disconnected sites. Under these conditions, the learning protocol $\mu(t)$ introduces a probabilistic weight flip, whose initial condition is given by the steady-state amplitude of the computational degrees of freedom. Figure~\ref{fig:lattice}d,e illustrates the training of a metamaterial to match a specific signal transmissivity. At every training iteration, the expected value of the output energy gets closer to the target value, while the variance between different runs decreases.

\subsection*{Learning the Iris dataset}\label{Results}
\begin{figure}[b!]
\includegraphics[width = \columnwidth]{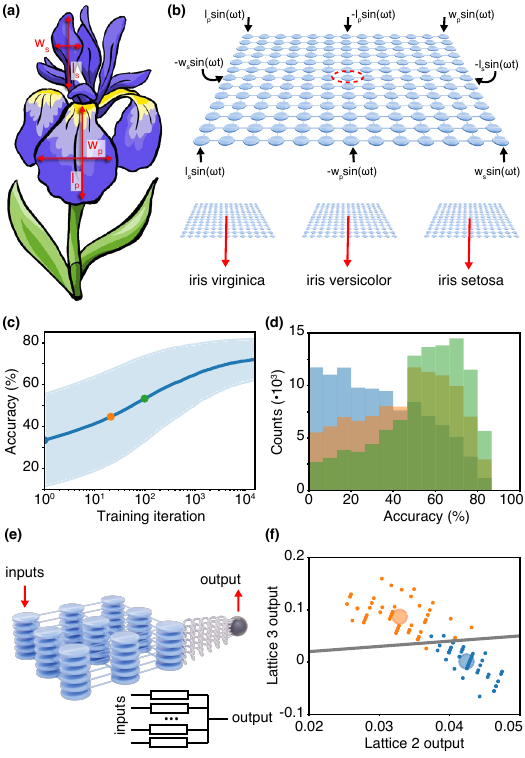}
\caption{Iris flower classification using a multifield coherent Ising machine. \textbf{(a)} Flower features contained in the Iris dataset. \textbf{(b)} The features are injected into the multifield coherent Ising machine, by encoding them in the amplitude of harmonic excitations at the computational frequency $\omega_c$. The output is taken at the central site. Positive and negative copies of the signals are applied, as the lattice cannot perform subtractions. The full model consists of three multifield Coherent Ising Machines, corresponding to each of the model classes. The machines learn to produce a high amplitude when excited by a sample of their corresponding class. \textbf{(c)} Evolution of the mean classification accuracy during training. The shaded area corresponds to one standard deviation. \textbf{(d)} Histograms of the classification accuracy computed on an untrained lattice (blue), and after 20 (orange) and 98 (green) training iterations. The training times corresponding to the histograms are indicated as solid dots in panel (c).}
\label{fig:iris} 
\vspace{-15  pt}
\end{figure}

We illustrate the capabilities of the proposed system on the Iris classification test~\cite{fisher1936use}. The dataset contains measurements of the petal length $l_p$, petal width $w_p$, sepal length $l_s$ and sepal width $w_s$ (Fig.~\ref{fig:iris}a) for a set of 150 flowers belonging to the species \emph{iris setosa}, \emph{iris virginica}, and \emph{iris versicolor}. The classifier consists of a system of three 13x13 lattices; the lattice with the highest output amplitude, measured at the central site, will be the inferred flower species. We encode the geometric features ($l_p$, $w_p$, $l_s$, $w_s$) in the amplitudes of a set of harmonic excitation forces acting on each lattice. These signals are applied at sites through the boundary of the lattice (Fig.~\ref{fig:iris}b), acting on both free $\psi_x$ and clamped $\psi_y$ computational modes. The input and output locations are selected so that the output is at an approximately equal distance from all inputs. Since the lattices cannot produce large negative transmissivities (Fig.~\ref{fig:lattice}b), we apply both positive and negative copies of each feature signal. Alternative interaction potentials can actually produce negative transmissivities, as is discussed in the appendix. 

The training process starts by initializing each of the three lattices with a random weight configuration. Then, for every learning iteration, we subject all lattices to the forces encoding the features of a randomly chosen flower.  We clamp the output site $\psi_{y,center}$ to $55$ for the lattice corresponding to the current type of flower, and to $-35$ for the lattices corresponding to the other flowers (these values have been chosen empirically to fall within the range of expressible values). Finally, we apply a weight update protocol to all lattices, consisting of setting the coupling $\lambda$ to zero, and changing $\mu(t)$ following a Gaussian pulse. This process is repeated for 2000 iterations.

We observe that the system starts from an accuracy of $33.2 \pm 22.1\,\%$ , as one would expect from random choice. As more training iterations are performed, the accuracy increases to $71.9\pm9.8\,\%$, providing a clear signature of physical learning. These results are remarkable as the system stores only 156 bits of information, contains only Kerr nonlinearities, and is directly trained by exposing it to the desired output. Using a more advanced training algorithm, where the learning process is stopped when the precision exceeds a threshold, we achieve $80.0 \pm 7.0\,\%$ accuracy, with the best-performing runs reaching $87.3\,\%$. Although these accuracy figures are much lower than traditional digital machine learning algorithms (methods such as k-nearest neighbors, support vector machines and decision trees achieve performances in the $93.0-97\,\%$ range), and to electronic implementations of learning rules (which can already reach values around $95\,\%$), they are remarkable in that they are achieved through the dynamical evolution of a nonlinear mass-spring network (subject to examples and simple clocking pulses). Our accuracies are however in line with similar 'unconventional learning' demonstrations, such as classifying with origami sheets (which reach best-values of $92\,\%$ on a simpler two-class flower test) or when using EEG signals (which achieve accuracies $80\,\%$). To confirm that our classification provides evidence of learning, and is not a result of random fluctuations, we perform statistical hypothesis testing in the Appendix.

Although in line with other unconventional learning results, the limited accuracy of our trained classifier merits further investigation. We identify two possible limiting factors: The binary nature of the weights in our system---which may reduce the model expressivity, or peculiarities of the training protocol. To separate the two effects, we construct a digitally-trained multiclass linear architecture using networks of $3x3$ resonator lattices, which are highly quantized. The model projects a feature vector $\vec{f}=(p_l, p_w, 1)$ onto a set of weight vectors corresponding to the three flower classes ${\vec{w}_{virginica}, \vec{w}_{versicolor}, \vec{w}_{setosa}}$. Every class is implemented using a network of three pairs of $3x3$ lattices connected to an output mass (Fig.~\ref{fig:iris}e), with each pair of lattices expressing the positive and negative values of a specific weight vector element. Using such architecture, as opposed to a large lattice, allows us to identify specific sub-lattices with model coefficients. We initialize the spin texture of every lattice to the closest value required by the linear model. When simulating the lattice dynamics, the model achieves an accuracy of $97.3\,\%$---the maximum possible for a linear classifier. This indicates that weight quantization cannot explain the observed performance degradation

Therefore, the low precision on the Iris dataset must originate in the training dynamics. We know from  Fig.~\ref{fig:}b, that lattices can be trained to arbitrary transmissivities within their range of expressivity. This works because spin changes stop taking place when the transmitted amplitude matches the target. However, when training on the Iris dataset, each sample causes the lattice to shift towards a different transmissivity (Fig.~\ref{fig:iris}f), preventing the lattice from settling into a stable value. In prior, high-accuracy, electronic realizations of contrastive learning~\cite{labDem}, the training process has not prescribed the output to a specific value, but instead \emph{nudged} incorrect answers towards the correct value. We hypothesize that such sophisticated training methods may enable the system to achieve higher accuracies---potentially comparable to electronic linear classifiers, by allowing it to settle into high-accuracy configurations even if the output does not match a specific transmitted amplitude. However, evaluating such more complex protocols falls outside the scope of this work, which focuses on control-free systems, i.e. where the input signals applied to the network are independent of its state.

\section*{Discussion}\label{Intro}
We have shown that a metamaterial composed of a network of parametrically-driven, multimodal resonators can learn to solve a classification task, when excited with example signals. The material performs its own weight update according to a learning rule, requiring only two simple drive clock signals $\lambda(t)$ and $\mu(t)$ in addition to the data and labels. However, the metamaterial does not require independent controllers adjusting each individual weight. The metamaterial learning dynamics is based on contrastive learning, which involves comparing two copies of the system---one left free, and one clamped to the desired output---and updating the model weights according to the difference between copies. However, unlike traditional contrastive learning, our model weights are binary and cannot be updated continuously. We circumvent this limitation by reformulating the learning rule as a probabilistic process, where the probability of a weight flip is a function of the mismatch between the two copies. This mismatch-dependent flip rate is reminiscent of evolutionary mechanisms in bacteria, where the mutation rate increases when the organism is under stress~\cite{bjedov2003stress, gutierrez2013beta, pribis2022stress}. Similarly, our system increases the weight mutation rate at sites where its clamped and free copies disagree.

In our model, learning emerges from the interplay of symmetry, multistability, and noise, features not typically associated with computation. This work therefore contributes to the growing evidence that such physical effects hold significant, untapped potential for information processing~\cite{aifer2024thermodynamic, melanson2025thermodynamic, harabi2023memristor, chen2025solving}. In fact, symmetry plays an essential role in the learning process: Clamping the $\psi_y$ breaks the symmetry between the two components of the computational field $\vec{\psi}$. The weight field, $\sigma$, then evolves to restore this broken symmetry over time. As symmetry is gradually restored, the free part of the system learns to mimic the clamped part, not by accessing the training data directly, but through persistent changes in the weight variable $\sigma$. This perspective suggests a novel symmetry-based design principle for self-learning materials, much like symmetry considerations guide the design of emergent condensed matter phenomena such as topological edge states~\cite{altland1997nonstandard}. The presence of multiple time scales is also an essential component. These are important in the brain~\cite{Masset2025} and can be replicated in artificial systems through parametric drive.

By providing a tractable physical model for emergent learning---analogous to the tight-binding Hamiltonians that describe topological insulators and superconductors---our work invites condensed matter-inspired questions about learning. For instance, one can now investigate the influence of disorder~\cite{zu2024fully, krushynska2022multi}, aging~\cite{dohare2024loss}, and phase transitions~\cite{kirkpatrick1994critical, amit1985spin, power2022grokking} on learning behavior. Although the present work is concerned with non-dimensional tight-binding models and thus no specific performance claims can be made, our model may inspire new directions to train machine learning models with physical computers, as the constituent oscillators can be realized with components that operate at optical frequencies~\cite{okawachi2020demonstration}, consume little energy~\cite{cupertino2024centimeter}, or are integrated with very high densities~\cite{zahedinejad2020two}. Our work also poses novel exciting fundamental questions, for example involving learning dynamics in discrete spin systems, architectural and scaling considerations, as well as addressing the experimental realization of such models.

\begin{acknowledgments}
We are thankful to Menachem Stern, Hermen-Jan Hupkes, Cyrill B\"osch, Henrik Wolf, Tena Dub\v{c}ek and Martin van Hecke for helpful discussions. The illustrations in Fig. 1, 3a and 4a,b have been provided by Laura Canil from Canil Visuals. 

Funded by the European Union. Views and opinions expressed are however those of the author(s) only and do not necessarily reflect those of the European Union or the European Research Council Executive Agency. Neither the European Union nor the granting authority can be held responsible for them. This work is supported by ERC grant 101040117 (INFOPASS). 
\end{acknowledgments}

\bibliography{manuscript}% Produces the bibliography vi

\appendix
\section{Determination of the parametric resonance regions}\label{Appendix:PR}
In the manuscript we discussed how detuning of the parametrically pumped Ising mode leads to conditional forgetting. Here, we provide an in-depth analysis of parametric resonance using a combination of Floquet theory and perturbation methods. Our main objectives are to derive the resonance curve shown in Fig. \ref{fig:flip}a and to show that detuning has the effect of shifting it. The relevant equation is that of the Ising mode (Eq. \ref{eqn:Ising}) without interactions and noise, which can be cast in the following form:
\begin{equation}
\ddot{x} + \mu\dot{x} + \left(\delta + \epsilon\sin(t)\right)x + \gamma x^3 = 0,\label{eqn:math_PO}
\end{equation}
where $\delta$ represents the squared ratio between the mode's natural frequency and the pumping frequency and $\epsilon$ represents the driving amplitude. Resonance occurs because the pumping effectively serves as a negative damping. Growing oscillations are detuned by the Duffing term, restricting the steady state amplitude. So, for fixed damping $\mu$, the relevant parameters that control whether or not resonance occurs are $\delta$ and $\epsilon$. To derive the resonance curve in the $\delta-\epsilon$ plane we treat the damping and Duffing terms as small perturbations to the well-known and extensively studied \emph{Mathieu equation}:
\begin{equation}
    \ddot{x} + \left(\delta + \epsilon\sin(t)\right)x  = 0,
\end{equation}
for which the resonance regions can be found using Floquet theory. They form tongues emanating from the points $\delta = n^2/4$ along the $\delta-$axis in what is known as the Ince-Strutt diagram~\cite{Mathieu_eq}. For the purposes of this paper, we consider the resonance region around the so-called \emph{parametric resonance condition}, which holds when the driving frequency is twice the mode's natural frequency, or equivalently when $\delta = 1/4$. We can use perturbation theory to find an approximate expression for this resonance curve in the case where $\mu, \gamma > 0$, following an approach adapted from~\cite{Mathieu_eq} known as the \emph{method of multiple scales}. Firstly, we treat the pumping amplitude $\epsilon$ as a small perturbation, and rescale $\mu$ and $\gamma$ to be of the same order, $\mu =: \epsilon \tilde{\mu}$ and $\gamma =: \epsilon \tilde{\gamma}$. Next, we expand the resonance curve in powers of $\epsilon$, $\delta = 1/4 + \delta_1\epsilon + \dots$ and define two independent timescales; a fast one $\xi := t$ and a slow one $\eta := \epsilon t$. Because the timescales are assumed to be independent, the full time derivative with respect to $t$ splits into partial derivatives with respect to $\xi$ and $\eta$ as follows:
\[\dv{}{t} = \pdv{}{\xi} + \epsilon \pdv{}{\eta}\]
Substituting everything into Eq.~\ref{eqn:math_PO} and ignoring terms of second order or higher in $\epsilon$, leaves us with: 
\begin{equation}
\frac{\partial^2x}{\partial \xi^2} + 2\epsilon\frac{\partial^2x}{\partial \eta\partial \xi} + \epsilon\tilde{\mu} \frac{\partial x}{\partial \xi} + \left(\frac{1}{4} + \epsilon \delta_1 + \epsilon \sin(\xi)\right)x + \epsilon \tilde{\gamma} x^3 = O(\epsilon^2). \label{eqn:pertubed_math_PO}
\end{equation}
Expanding the solution $x$ in powers of $\epsilon$, $x = x_0 + x_1\epsilon + \dots$ and collecting terms of like order yields the following two equations (one for the zeroth and one for the first order):
\begin{align}
   \frac{\partial^2x_0}{\partial \xi^2} + \frac{1}{4}x_0 &= 0  \\ 
   \frac{\partial^2x_1}{\partial \xi^2} + \frac{1}{4}x_1 &= -2\frac{\partial^2x_0}{\partial \eta\partial \xi} - \tilde{\mu} \frac{\partial x_0}{\partial \xi} - \delta_1x_0 - \sin(\xi)x_0 - \tilde{\gamma} x_0^3. 
\end{align}
From which we find the general solution for the zeroth order: 
\begin{equation}
x_0 = A(\eta)\cos\left(\frac{\xi}{2}\right) + B(\eta)\sin\left(\frac{\xi}{2}\right).  
\end{equation}
Substituting this in the equation for the first order, and collecting resonance terms on the right hand side we find:
\begin{align*}
 \frac{\partial^2x_1}{\partial \xi^2} + \frac{1}{4}x_1 &=  \sin\left(\frac{\xi}{2}\right)\left(A' + A\frac{\tilde{\mu}}{2} - B\delta_1\right)\\
 &- \cos\left(\frac{\xi}{2}\right)\left(B' + B\frac{\tilde{\mu}}{2} + A\delta_1\right)\\
    & -\sin\left(\frac{\xi}{2}\right)\left(\frac{3}{4}\tilde{\gamma} B^3 + \frac{3}{4}\tilde{\gamma} A^2 B + \frac{1}{2}A\right) \\
    &- \cos\left(\frac{\xi}{2}\right)\left(\frac{3}{4}\tilde{\gamma} A^3 + \frac{3}{4}\tilde{\gamma} AB^2 + \frac{1}{2}B\right)\\
& + \textit{off-resonance terms}.
\end{align*}
The terms on the left hand side that are on resonance are known as \emph{secular terms}~\cite{pertubation_methods}, and they cause the first order $x_1$ to grow without bound. So, in order for the perturbation assumption to be valid, these terms need to vanish. Setting the coefficients of the secular terms to zero leaves us with the following equations describing the (slow) dynamics of the amplitude and phase modulating the (fast) zeroth order oscillations $x_0$:
\begin{align*}
\frac{dA}{d\eta} &= \frac{1}{2}A\left(1 -\tilde{\mu}\right) + \delta_1B + \frac{3}{4}\tilde{\gamma} B\left(A^2 + B^2\right)\\ 
\frac{dB}{d\eta} &= -\frac{1}{2}B\left(1 +\tilde{\mu}\right) - \delta_1A - \frac{3}{4}\tilde{\gamma} A\left(A^2 + B^2\right). 
\end{align*}
Note that the amplitude of the zeroth order oscillations is given by the length of the vector $(A, B)$. So, when the origin of the dynamical system above is stable, small oscillations of the parametrically driven mode die out, and when it is unstable, small oscillations are amplified. Therefore, the parametric resonance curve is defined by the parameter values at which the linearized system transitions from having no eigenvalues with positive real part, to having at least one. The eigenvalues of the linearized system are given by:
\[\lambda_\pm = -\frac{\tilde{\mu}}{2} \pm \sqrt{\frac{1}{4}-\delta_1^2}.\]
From this we conclude that the origin transitions from unstable to asymptotically stable when $\lambda_- = 0$. With some rearrangements, substituting back in $\delta$ and $\mu = \tilde{\mu}/\epsilon$, we find the following equation for the resonance curve in the $\delta-\epsilon$ plane:
\begin{equation*}
\left(\delta - \frac{1}{4}\right)^2= \frac{1}{4}(\epsilon^2 - \mu^2).
\end{equation*}
Finally, we can re-dimensionalize the expression by introducing the driving frequency $\omega_d$, which in Eq. \ref{eqn:Ising} is set to $2\omega_l$ in accordance with the parametric resonance condition: 
\begin{equation*}
\alpha^2 = \frac{\omega_d^2}{\omega_l^2Q_l^2} + \left(\frac{\omega_d^2}{2\omega_l^2} - 2\right)^2. \label{stability_curve_phys}   
\end{equation*}
This is the expression used to determine the region of self-resonance in Fig.~\ref{fig:flip}. It should be noted that its value determines the minimum pumping that leads to self-resonance. Numerical simulations are conducted at twice this value, to ensure sufficient stability of the self-resonant state.

\section{Numerical methods}\label{Appendix:NumMeth}
We solve the stochastic differential equations (SDEs) using a Splitting Path Runge-Kutta solver (SPaRK), while for the deterministic simulations (four dots in Fig.~\ref{fig:lattice}b), we use a Dormand-Prince algorithm with order 7-8. Both solvers are implemented in Python using the Diffrax library. We choose a time step for a minimum resolution of 50 points per period of oscillation.

\subsection{Simulation of the learning process}
We simulate the learning process by computing the equilibrium amplitudes of $\vec{\psi}$ using an effective, linearized mass-spring model for the computational degrees of freedom. The effective model is calculated around the resonance frequency $\omega_c$. The coupling springs are given by the linearized version of $Eq.~\ref{eqn:interactionpotential}$, while the diagonal terms are governed by the damping---in resonance, local inertia cancels with local stiffness---and take the form $i\omega_c^2/Q_c$. Once the computational field has been calculated using the linearized model, we flip the weights according to the single-site flip probability (Eq.~\ref{eqn:switchprobability}), based on the difference between clamped and free fields. The single-site flip probability provides an accurate result as we switch off the inter-site potential during coupling, thus reducing the model to a set of independent elements. 

For the learning examples of Fig.~\ref{fig:lattice} and Fig.~\ref{fig:iris}, we use the parameters $k_{high}=10\omega_c^2/Q_c$, $4k_{low}=k_{high}$, $\beta=0.273$ and $\psi_T=55$. In Fig.~\ref{fig:lattice}, we use the flip probability parameters $\beta=750$ and $\psi_T=0.16$, while in Fig.~\ref{fig:iris}, the parameters are $\beta=0.273$ and $\psi_T=55$.

\subsubsection{Simulation of the clamped boundary conditions}
To impose the boundary conditions, we divide the degrees of freedom into internal (subscript $I$) and output (subscript $o$) components. We write the system $\mathbf{D}\mathbf{u} = \mathbf{f}$ in block form as:
\begin{equation}
\begin{bmatrix}
\mathbf{D}_{II} & \mathbf{d}_{Io} \\
\mathbf{d}_{oI}^T & d_{oo}
\end{bmatrix}
\begin{bmatrix}
\mathbf{u}_I \\
u_o
\end{bmatrix}
=
\begin{bmatrix}
\mathbf{f}_I \\
f_o
\end{bmatrix}
\end{equation}
where $u_o$ is the prescribed output degree of freedom, $f_o$ is the constraint force, and $F_I$ is the external force (which inputs the data into the system). From the first block row, we bring the prescribed output to the right-hand side to obtain the reduced system $\mathbf{D}_{II}\mathbf{u}_I = \mathbf{f}_I - \mathbf{d}_{Io}u_o$, which we solve for the unknown internal DOFs $\mathbf{u}_I$. Note that in the reduced system equation, it is not necessary to explicitly calculate the constraint force $f_o$ to simulate the system dynamics.

In an experimental setting, the boundary conditions could be applied by connecting the output degree of freedom to a very large mass, or by applying external feedback.

\subsection{Disorder and defect site tolerance}
\begin{figure}[h!]
\includegraphics[width = \columnwidth]{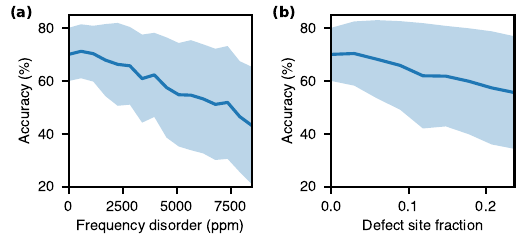}
\caption{Sensitivity to disorder. \textbf{(a)} Classification accuracy on the Iris dataset as a function of the resonator frequency disorder. \textbf{(b)} Classification accuracy on the Iris dataset as a function of the fraction of defective sites. Both results have been obtained by averaging 16000 simulations. The shaded area corresponds to one standard deviation.}
\label{fig:disorder} 
%\vspace{-15  pt}
\end{figure}

For our proposed learning mechanism to be accessible in a future experiment, it must be able to survive under the presence of realistic fabrication variability. To demonstrate that this is indeed the case, we simulated how the learning performance on the Iris task is affected by the presence of two different types of imperfections, frequency disorder and damaged resonators. For each type of imperfection, we conducted 16000 simulations, each of them with a different random perturbation (either a different frequency perturbation or a different choice of stuck resonators). We modelled the frequency disorder by adding a random, independent Gaussian increment to the frequency of each resonator (Fig.~\ref{fig:disorder}a), and simulated stuck resonators by prescribing their associated spin state to zero (Fig.~\ref{fig:disorder}b).

Simulations reveal that learning is not appreciably impaired by disorders below 250 ppm (parts-per-million) or by a fraction of defective resonators below 2.5\%. These numbers fall within the accessible experimental range: Resonators with frequency disorders of a few tens of ppm or better are routinely attainable in photonics~\cite{opticalTolerance} and mechanics~\cite{mechanicalTolerance}; similarly, defect free arrays of 50-100 devices are also routinely produced in both domains~\cite{opticalYield, memsYield}. These results provide strong evidence that, while the experimental realization of self-learning oscillator networks represents a significant experimental endeavor, it does not rely on unrealistic fabrication tolerances.

\subsection{Operation under a square learning protocol}
\begin{figure}[h!]
\includegraphics[width = \columnwidth]{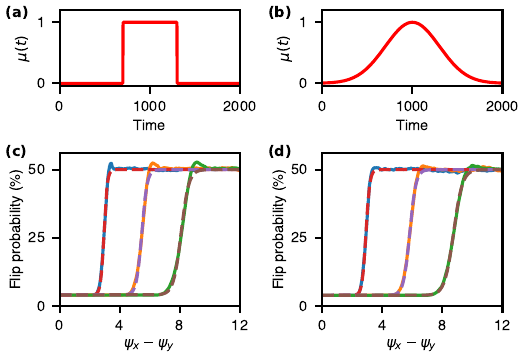}
\caption{Single-site flip probability with different pulse shapes.  \textbf{(a)} Numerical simulation of the transition probability when $\mu(t)$ varies according to a square pulse.  \textbf{(b)} Numerical simulation of the switch probability when $\mu(t)$ follows a Gaussian pulse, reproduced from the main text for convenience.  }
\label{fig:squarepulse} 
\vspace{-15  pt}
\end{figure}

Throughout the paper, we computed the transition probabilities assuming the learning pulse $\mu(t)$ followed a Gaussian function. Generating such pulses may require additional circuitry, therefore, it is worth examining whether a simpler function, such as a square pulse, can achieve the same results. Figure~\ref{fig:squarepulse}  compares the transition probabilities obtained by switching the learning potential according to a square function (Fig.~\ref{fig:squarepulse}a), 
\begin{align*}
\mu(t) &=
\begin{cases}
\mu_m, & \text{if } |t - t_0| < \Delta_t,\\[4pt]
0,     & \text{otherwise},
\end{cases}
\end{align*} with the Gaussian pulse results obtained in the paper. The simulations reveal that both pulses achieve similar sigmoid-like probability transitions, indicating that it suffices to switch on the learning potential when a data sample is applied.

\section{Realization of the interaction potential with nonlinear springs}\label{Appendix:NonlinearSprings}
The spin-dependent part of potential from Eq.~\ref{eqn:interactionpotential} can be approximated by a set of cubic springs, governed by the Hamiltonian:
\begin{equation}
H_{I,ij}=\lambda'(t)\sum{\alpha_kd^4_{v_{k,i}-u_{k,j}}},
\label{eqn:interactionpotentialSpring}
\end{equation}
where $d_{v_{k,i}-u_{k,j}}=v_{k,i}-u_{k,j}$ is a difference  between the field value $v_{k,i}$ at drum $i$ and the field value $u_{k,i}$ at drum $j$. Field values are defined as a linear combination of modes, $v_{k,i}=\vec{\kappa}^T_k \cdot (\psi_{x, i}, \psi_{y, i}, \sigma_{i})$ and $u_{k,j}=\vec{\kappa}^T_k{}' \cdot (\psi_{x, j}, \psi_{y, j}, \sigma_{j})$ respectively.  The coefficients $\vec{\kappa}_k$, $\vec{\kappa}_k{}'$ and value $\alpha_k$ have been calculated to induce a spin-dependent spring between sites, while compensating for spin-dependent local frequency shifts in the computational and long-term degrees of freedom. These coefficients are provided in Table \ref{tab:coefficients}. Although our work concerns a tight-binding model, a potential experimental realization might involve a mechanical resonator network, where each of the coupling terms in Eq. \ref{eqn:interactionpotential} corresponds to a cubic spring connecting a point $\vec{r}_{1,k}$ in resonator $i$ where the tight-binding basis functions $\phi_{\sigma/\psi_x/\psi_y}(\vec{r}_{1,k})$ take the values $\kappa_{1/2/3,k}$ with a point $\vec{r}_{2,k}$ of resonator $j$ where the tight-binding eigenfunctions $\phi_{\sigma/\psi_x/\psi_y}(\vec{r}_{1,k})$ take the value $\kappa_{1/2/3,k}'$. 

\begin{table}[h]
    \centering
    \begin{tabular}{|c|c|c|}
    \hline
    $\alpha_k$ & $\kappa_k$ & $\kappa_k'$ \\
    \hline
    1 & $(1,0,1)$ & $(1,0,1)$ \\
    1 & $(0,1,1)$ & $(0,1,1)$ \\
    \hline
    -1 & $(1,0,1)$ & $(0,0,1)$ \\
    -1 & $(0,0,1)$ & $(1,0,1)$ \\
    -1  & $(0,1,1)$ & $(0,0,1)$ \\
    -1  & $(0,0,1)$ & $(0,1,1)$ \\
    \hline
    2  & $(0,0,1)$ & $(0,0,1)$ \\
    \hline
    \end{tabular}
    \caption{Coefficients for the inter-site interaction potential. }
    \label{tab:coefficients}
\end{table}

Of the seven interaction potential terms in Table~\ref{tab:coefficients}, the first two introduce the spin-dependent inter-site interactions between modes $\psi_x-\psi_x'$ and $\psi_y-\psi_y'$ respectively; the next four terms compensate for spin-dependent changes in the natural frequency of the computational degrees of freedom, while the last term compensates for spin-dependent changes in the Ising degrees of freedom. These extra terms are required to cancel the contribution of cross-products in Eq.~\ref{eqn:interactionpotentialSpring}, that do not exist in Eq.~\ref{eqn:interactionpotential}. Only the first two terms determine the flow of information, as they couple computational degrees of freedom between sites. In absence of the compensating terms, the system may still be capable of learning. However, in this case, the output phase becomes dependent on the spin texture, which complicates the learning process.

The values in Table~\ref{tab:coefficients} have been determined by expanding Eq.~\ref{eqn:interactionpotentialSpring} and choosing the coefficients to cancel on-resonance terms other than those in Eq.~\ref{eqn:interactionpotential}.

 We expect that these results will motivate experimental realizations of the model, for example in metamaterial or optomechanical systems, in the same way that the emergence of topological tight-binding Hamiltonians~\cite{benalcazar2017quantized} motivated the realization of metamaterial topological insulators~\cite{serra2018observation, peterson2018quantized, matlack2018designing}. In addition, this work poses a broad spectrum of novel questions exploring the realization of alternative learning potentials, learning rules and network architectures; the study of systems with more than two memory scales; as well as of the kinetics of the learning process.
 
 \subsubsection{Simulation of learning under a spin-dependent local potential}
 \begin{figure}
 \includegraphics[width = \columnwidth]{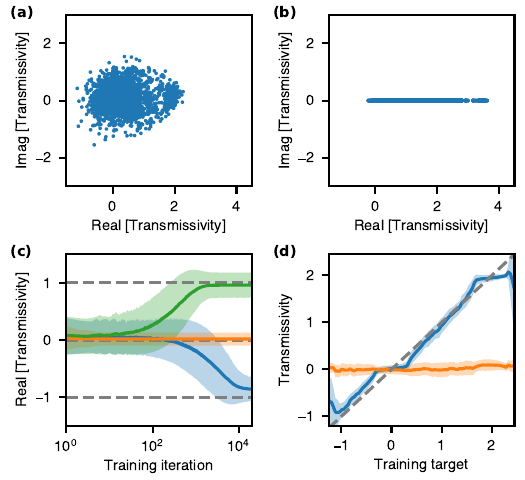}
 \caption{Training of a 4x4 lattice with uncompensated interactions. \textbf{(a, b)} Complex transmissivity values for each of the $2^{16}$ possible spin configurations in the absence (a) and presence (b) of compensating terms. \textbf{(c)} Evolution of the real part of the transmissivity as the lattice is trained towards target transmissivities of $-1$ (blue), $0$ (orange) and $1$(green). \textbf{(d)} Final value of the real (blue) and imaginary (orange) parts of the transmissivity as the lattice is trained towards different real targets.}
 \label{fig:variableLocalPotential} 
 \end{figure}
In Table~\ref{tab:coefficients}, only two of the coupling terms are essential for learning. The other five are introduced to compensate for the spin-dependent shifts on the natural frequencies of the various modes, i.e., the term ensures that the frequency of the computational modes $\psi_{x/y}$ is independent of the spin configuration. In Figure~\ref{fig:variableLocalPotential}, we examine the learning dynamics when such compensation is not in place, and the local potential $\omega_{l}$ is dependent on the spin texture. Under these conditions, the output signal can take an arbitrary complex phase (Fig.~\ref{fig:variableLocalPotential}a), differing from the original scenario where the transmitted amplitude had a fixed phase relation with the lattice excitation (Fig.~\ref{fig:variableLocalPotential}b). Simulating the learning dynamics as the system is trained towards a fixed reference transmissivity reveals that the system is still able to learn (Fig.~\ref{fig:variableLocalPotential}c), although the number of iterations that it takes to converge to the correct transmittance increases by a significant amount. We attribute this increase to the fact that the system presents fewer spin configurations that result in the required transmissivity value---as transmissivity values are spread around the complex plane; therefore, it takes longer to settle in the correct configuration. Remarkably, this system can be trained to negative transmissivities (Fig.~\ref{fig:variableLocalPotential}c, d), potentially relaxing the need for inverse inputs in the Iris dataset test.

\begin{figure}
 \includegraphics[width = \columnwidth]{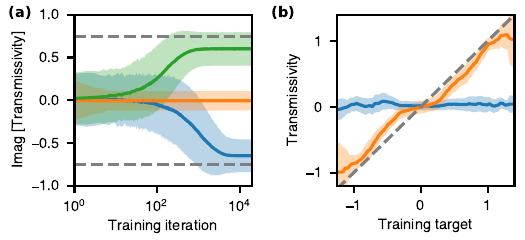}
 \caption{Training a 4x4 lattice to produce phase-shifted signals. \textbf{(a)}  Evolution of the imaginary part of the transmissivity as the lattice is trained towards targets of $-0.75i$ (blue), $0$ (orange) and $0.75i$ (green). \textbf{(b)} Final value of the real (blue) and imaginary (orange) parts of the transmissivity as the lattice is trained towards different imaginary targets. }
 \label{fig:trainToImaginaryTransmissivity} 
 \end{figure}

These results motivate us to explore whether the system can be trained beyond signal inversion, to produce signals that are phase-shifted by 90 degrees. We observe that this is indeed the case (Fig.~\ref{fig:trainToImaginaryTransmissivity}a, b). These results indicate that, although presenting slower learning dynamics, removing the compensating terms from Table~\ref{tab:coefficients} also increases expressivity.
 
  \subsection{Selection of resonator eigenmodes}
\begin{figure}
\includegraphics[width = \columnwidth]{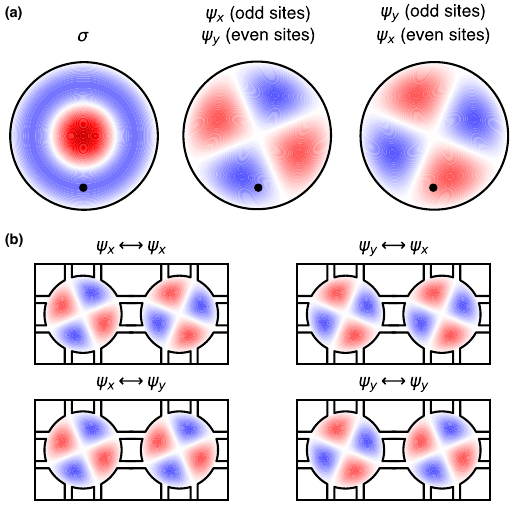}
\caption{Mapping model degrees of freedom to resonator eigenmodes. \textbf{a} Possible mode selection for the weight ($\sigma$) and computational ($\psi_x$ and $\psi_y$) degrees of freedom. Note that for the computational modes, the assignment between variables and local modes (gauge) varies between even and odd sites. The black dot indicates a region where the computational modes have opposing sign, and overlap with the weight mode. There, the nonlinear learning potential $H_L$ would be applied. \textbf{b} By placing the nonlinear coupling elements at the zeros of the computational modes, we can realize the simplified coupling interactions. }
\label{fig:modeprofiles} 
\end{figure}

Here, we will investigate how a self-learning metamaterial could be constructed, following the simplified model discussed in the previous section. To realize a metamaterial with the right interactions, we need to identify mode pairs with alternating zeros at the boundary. These zeros allow for the placement of the coupler beams that realize individual interaction terms. Remarkably, the eigenpair with angular mode number $\ell = 2$ (Fig.~\ref{fig:modeprofiles}a) of a circularly-symmetric resonator naturally lends itself to realizing the required interaction---as both modes in the pair expose regions of zero field at the boundary with all four neighbors, and the coupler beams can be placed (Fig.~\ref{fig:modeprofiles}b). Because these properties originate in the symmetry characteristics of the mode, and thus do not rely on specific physics of the system (photonics, MEMS, microwave...), this mode selection provides a path for the realizaton of self-learning materials in diverse platforms.

In Fig.~\ref{fig:modeprofiles}a we have also highlighted the location for the application of the learning potential $H_{l}$, which couples the difference between computational modes $\psi_{x/y}$ to the learning mode $\sigma$. By selecting a learning mode with radial mode number $n>0$, we also construct regions where only one mode is nonzero, as well as regions where only two of the modes is nonzero (for any pair of modes). This allows the design to incorporate tuning electrodes for specific modes, as well as couplers for specific pairs. 

While these modes correspond to free-standing drums, in weakly-coupled resonators the mode shape can be assumed to remain constant up to first order of perturbation theory~\cite{matlack2018designing, fard2025embodying}, and differences can be compensated systematically by editing the resonator geometry.

\subsection{Testing the hypothesis that oscillator lattices can learn}
This article provides evidence for the hypothesis that an oscillator network can learn to solve a classification problem. Because the site flips are probabilistic, there is always a possibility that the observed improvement in classification accuracy is a random occurrence and does not represent genuine learning. To determine the strength of the results, we perform a paired t-test, which allows us to determine the p-value, the probability to randomly obtain (null hypothesis) a final distribution at least as extreme as the result from the simulations. The p-value is calculated from a t-value, given by
\begin{equation}
t=\frac{<d>\sqrt n}{\sigma(d)},
\label{eqn:tvalue}
\end{equation}
Where $d_i$ is the difference between initial and final classification accuracies, and $\sigma(d)$ represents its sample standard deviation. 
Evaluating Eq.~\ref{eqn:tvalue} for the initial and final data from Fig.~\ref{fig:iris}c results in a t-value of 180.01, with 12800 samples. 

For very large number of samples, the t-distribution is approximately Gaussian. Hence, the corresponding p-value is given by
\begin{equation}
p\approx\frac{1}{t\sqrt(2\pi)}e^{-t^2/2},
\label{eqn:tdtsrlimit}
\end{equation}
These results indicate that the probability of randomly obtaining a performance improvement as extreme as the one from Fig.~\ref{fig:iris}c is on the order of $e^{-16200}$. These results provide very strong evidence for the presence of learning: If the null hypothesis were true, the probability to obtain such an improvement would be approximately $10^{-7035}$.

\end{document}